# A MADM method for network selection in heterogeneous wireless networks

**Fayssal Bendaoud[1], Fedoua Didi[2], Marwen Abdennebi[3]**

[1,2]LRIT laboratory, University of Tlemcen, ALGERIA
[3]L2TI laboratory, University of Paris 13; FRANCE

**ABSTRACT**

The coexistence of different Radio Access Technologies (RATs) in the same area has enabled the researchers to get profit from the available networks by the selection of the best RAT at each moment to satisfy the user requirements. The challenge is to achieve the Always Best Connected (ABC) concept; the main issue is the automatic choice of the suitable Radio Access Technology (RAT) from the list of the available RATs. This decision is called the network selection (NS).

In this paper, we propose a modified Simple Additive Weigh (modified-SAW) function to deal with the drawbacks of the existing solutions. Indeed, the existing Multiple Attribute Decision Making (MADM) methods suffer mainly from the famous problem of rank reversal once an alternative is added or removed, other problems occur in the legacy MADMs. We modify the SAW method intelligently and we use it to solve the NS problem. Finally, we compare the performance of our solution with the previous works in different scenarios; the simulations show that our proposal outperforms the other existing methods.

## 1. INTRODUCTION

Up to the third generation network, the radio access network was mainly homogeneous and the user has to connect to only one RAT. After that, the development of the network technologies has led to an impressive growth of the internet applications and services, and an increasing in the mobile user's industry. People now are equipped with Smartphones try to achieve for the ABC concept. It is obvious that no single network technology can sustain that, therefore, it was necessary to change the whole design, i.e. switching from the homogeneous systems to the heterogeneous systems.

Nowadays, we have a variety of RATs, the WLAN basically IEEE802.11, UMTS, HSPA and the LTE. This variety constitutes the elements of the heterogeneous environment [1]. The heterogeneous system allows mobile users to choose a RAT among a list of RATs based on several criteria, this choice is called network selection and this is the scope of this paper.

The network selection procedure consists of the dynamic and the automatic selecting the best available network among the available RATs. The best network may differ from one user to another due to the different criteria involved cost, Quality of service (QoS), energy consumed, etc. The network selection in a heterogeneous environment can be reported to the MADM problems since it involves a huge number of criteria. The MADM approach has been widely used to solve the NS problem [2], [3], [4]. Other methods like fuzzy logic and game theory [5], [6], [7] have also been used to solve the network selection.

Our study seeks to find the best network, in addition, we give the users the other alternatives in the case when the best network is unreachable, i.e., we give the list of the ranked networks to select one of them according to the network load. In this paper, we try to bring out our solution based on the modified-SAW and we make a comparison with the legacy MADM methods.

The rest of the paper is organised as follows Section 2 presents the problem formulation and the related works, in Section 3 we give the mathematical description of the proposed method, in Section 4, the simulations are presented and the performance analysis is provided, finally a conclusion and perspectives are given in section 5.

## 2. NETWORK SELECTION
### 2.1. Problem context

In the Next Generation Network, the heterogeneous wireless access is a promising feature in which the users are sufficiently flexible to select the most appropriate network according to their needs. In these circumstances, the network selection has an important task for the smooth functioning of the whole communication system. Indeed, the NS process consists of switching between RATs to serve the user with the best network [6]. So, when a user with a multi-mode terminal discovers the existence of various RATs within the same area *Fig.1* he should be able to select the best network to get the desired service.

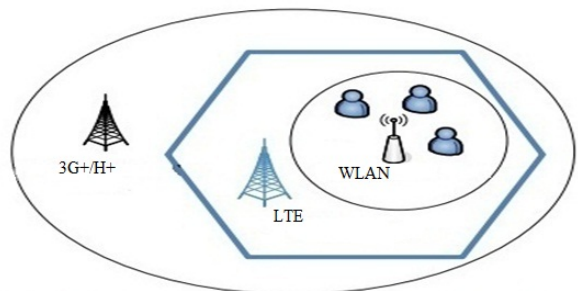

**Fig.1**: *Heterogeneous wireless environment [6].*

The different RATs provide different characteristics in terms of delay, jitter, throughput and packet loss rate. For this reason, in the context of heterogeneous systems, selecting the best RAT is a hard task. Many parameters influence the decision process of the best RAT [8], the battery level status level, the energy required for the requested services, the Signal to Interference plus Noise ratio received (SINR), the cost to pay, the bandwidth required, the user preferences, QoS, etc.
The NS procedure is the decisive part of the vertical handover process (VHO), it can be either centralised (network-centric) or decentralised (user-centric).

For the centralised approach, the operator controls the whole process and makes the decisions, the users obey these decisions and execute them, this strategy cannot be used in the case of multiple operators.

For the user-centric approach, users make decisions by themselves; Nowadays, almost all operators offer 3G and 4G radio access and even the Wi-Fi connections, so, the centralised approach is suitable to use. The network selection procedure consists of the following parts:

*Monitoring step*: the objective of this step is to discover the available RAT, collect the network radio conditions, and the other characteristics of the RATs.

*Decision step:* the network selection decision is started. The choice of the best network is based on the monitoring process. At this stage, a decision algorithm is used to rank the different RATs.

*Execution step:* consists on connecting to the target RAT.

**2.2. Related work**

Several works in the literature treat the NS problem, these works focus on the optimisation of the network selection decision for the users in order to support many services with the best QoS in a smooth manner.

In [9] the authors have used the Simple Additive Method SAW to get a ranked list of networks, while in [10] the authors make a mix between game theory and SAW method. The main benefits of SAW are the simplicity and the low complexity, it has two major drawbacks, first: a parameter can be outweighed by another one, second: the rank reversal phenomenon that represents a problem of the entire MADM approach.

The MADM methods are widely used to solve the network selection problem; this is due to the multiple criteria nature of this problem. The drawbacks of these methods are summarised in the following:
- The MADM methods don't have the same performance toward the different services (VoIP, Video Calls, web browsing), indeed, these methods have a good performance for the VoIP service and a bad performance for the best effort services. This instability involves great problems because users use different services.
- MADM methods suffer from the problem of ranking abnormality, i.e. it is a phenomenon that occurs in the MDAM methods when an exact replica or a copy of an alternative was introduced or eliminated, authors in [16] has shown that the rank reversal problem occurs in most of the well-known MADM methods, this problem has been addressed in other works [17], [18] by modifying methods, but the original versions of MADM methods suffer from the rank reversal phenomenon.

**2.3. MADM methods: Mathematical description**

The MADM approach is a famous mathematical approach in the context of preferential decisions; it treats problems involving many decision criteria and many alternatives. This branch of decision making is widely used in various fields such as the economy sector [19], [11], [9].

This approach MADM is very adapted to the network selection problem because of the multi-criteria nature of the NS problem [20]. In [21] the authors present the basics of this approach:

*Alternatives:* represent the different choices of actions available to the decision maker. The set of alternatives is assumed to be finite. In the NS scenario, the alternatives are the different RATs.

*Set of criteria:* called also goals, it represents the attributes used in the decision-making process. For our NS scenario, criteria are throughput, jitter, delay, cost, etc.

*Weights:* it means the importance of the criterion in the decision process, the sum of the weight is equal to one. Different methods are used to determine the weight values.

Finally, we get a decision matrix representing the system, where the columns are the criteria and the lines are the alternatives. Several methods use the MADM approach philosophy have been proposed in this context, such as Simple Additive Weighted SAW, Technique for Order Preference by Similarity to Ideal Solution TOPSIS, Weighted Product Model WPM, Analytic Hierarchy Process AHP, Grey relational analysis GRA, etc [11]. In the following, we give a general description of these algorithms.

- *Simple Additive Weight SAW:*

SAW is a well-known method for the case of multiple criteria systems, it assumes that data treated have the same unit, so to get a comparable scale among parameters, it is mandatory to normalise the data for each parameter [11], [9], [22]. Finally, the alternative with the highest value is selected. The mathematical formulation of SAW is:

$$R_{saw} = \sum_{i=0}^{N}(w_j * r_{ij}) \qquad (1)$$

$R_{saw}$: is the value of each alternative.

$w_j$ is the weight value of the criterion j.

N is the number of the decision criteria.

$r_{ij} = \frac{m_{ij}}{\max(m_{ij})}$ is the normalised value of the criterion j and the alternative *i*. Then, the chosen alternative is the one with the maximum value.

**3. THE MODIFIED-SAW ALGORITHM**

We have noticed that SAW method is so simple compared to TOPSIS and WPM (WPM has two versions and are hard to program with higher complexity). We have implemented these methods (TOPSIS and WPM) and we were not satisfied with the ranking order, so, we decided to improve the simple one which is SAW method. The proposed solution assigns the user to the best network from the available networks at the present instant while the selected network reachable i.e., not loaded. This process is repeated at several instances until the user's call session is ended.

So, when a user claims for a particular service, it sends a request to the operator, the request contains some information like the service required and the battery level. The other parameters needed for the network selection procedure are gathered by the operator. Then, the operator triggers the process of ranking the available networks. The results will be forwarded to the user who selects the best available network.

The forwarded result contains the list of the networks ranked from the best network to the worst one. Obviously, the user will choose the network having the best rank while the network is not loaded. A loaded network normally cannot gets the best rank, simply because a loaded network has a higher delay and lesser throughput, thus, it just has a bad QoS performance, so it is quite unlikely to get the best-ranked network a loaded network.

In our description based on *Fig 2*, we have two sides: the mobile user that seeks for the best RATs and the operator side that triggers the ranking process of the available networks. The idea behind the objective function is simple. We formulate the system as a minimising function in which the lowest value for each criterion gives the best-ranking order for the networks. Consequently, it gives it the highest local gain. This process is repeated until all the criteria will be evaluated. Its representation is as follows: for each network '*i*', we do:

```
Data: The matrix mat [m][n] and the weighting vector
      w[], α = m
Result: Income vector for each network
for (j = 0; j < m; j + +) do
    for (i = 0; i < n; i + +) do
        | tab[i] = mat[i][j];
    end
    tabind = Sort(tab);
    for (k = 0; k < Tabind.length; k + +) do
        | income[tabind[k]] =
            income[tabind[k]] + (α − k) × w[k]
    end
    return income[];
end
```
**Algorithm 1**: Modified-SAW function

$$R_i = \sum_{j=1}^{m} income_{ij} \quad (4)$$
$$income_{ij} = (\alpha - k_{ij}) * w[j] \quad (5)$$
$$k_{ij} = \min(Vect_{ij}) \quad (6)$$

Where:
$\alpha$: is a fixed integer number equal to the number of alternatives.
$k_{ij}$ : it represents the rank order of the network *i* for the criterion *j*.
$Vect_{ij}$ : is the column vector from the matrix *mat* where *j* is fixed.
*w[j]*: is the weight associated with the criterion "*j*" for the alternative "*i*".
*i*: is the alternative and *j* is the criteria.
*mat[n][m]:* is the input matrix, it is represented by Table2.
We start by dividing the input matrix to a set of column vectors to get a group of vectors equal to the number of criteria.

For each vector, the networks are ranked according to their data values. Then, each network receives a local income equal to the mathematical multiplication between '$\alpha$' minus the rank value for this network and the weight value of the criterion. '$\alpha$' is a fixed value equal to the number of criteria. This process is repeated for the other criteria using equation (5).

The total income value is equal to the sum of all the local incomes for each network, see equation (4).
We use the weight value of each criterion to differentiate between the criteria as all the MADM methods do. It is well known that the delay and the Packet loss Ratio (PLR) criteria are the most important comparing to the other criteria. They represent the QoS parameters.

For each network '*i*', the best case is having a minimal value for the criterion '*j*' this means the highest local ranking, thus $k_{ij} = 0$ and the revenue is $R_{ij} = \alpha * w[j]$ in this study $\alpha = 5$ (the number of criteria).

The worst case is when the network has the maximum value for the criterion '*j*', i.e., $k_{ij} = \alpha - 1$, i.e., the revenue is $R_{ij} = w[j]$

In this study, we modify the use of this concept weight vector by associating this concept with the '$\alpha$' value. This modification has mainly two advantages:
- Avoid the situation in which a network that has a good value for a non-important criterion will have the same revenue as another network with a good value for an important criterion. This situation exists in the SAW method for example due to the use of the ordinary weight vector (without the modification proposed in this work).
- The weight concept is the representation of application requirements in the system and that's how we distinguish between applications because each one has specific requirements. VoIP requires a minimum time delay and PLR for the video application, in addition to the requirements of VoIP service, it demands also a good throughput, for the best-effort applications, they accept the existing conditions, but the cost criterion is so important. This information is transformed into digital values with the eigenvector method.

In our study, we use many parameters such as cost, energy consumption, average throughput achieved, average delay, average PLR and network load.

The modified-SAW function consists on the decomposition of the input matrix in column vectors, after that, it uses the *Sort function* to rank the networks of these vectors and return their indices ranked from the best network to the worst; the ranked indices are stocked in the "*Tabind*" vector. After that, we calculate the local income for each network using the equation (5). The process is repeated for all the vectors with by making the mathematical sum of all the local incomes and like this, the local incomes become total incomes when we treat all the criteria.

The values of Table 1, the bandwidth, delay and PLR are margin values obtained from simulations with NS3.

*Table 1 The matrix model*

|       | Bwidth | Delay (ms) | PLR (%) | Energy | Cost |
|-------|--------|------------|---------|--------|------|
| Wi-Fi | 1-11   | 100-150    | 0.2-3   | -      | 1    |
| 3G    | 1-14   | 25-50      | 0.2-3   | -      | 5    |
| LTE   | 1-100  | 60-100     | 0.2-3   | -      | 2    |

The energy consumption parameter is related to the level of the battery of the mobile device as well as the duration of the call session. Based on [24] the energy required to execute the user's request will be calculated for each RAT. The energy consumption parameter is set using equation (7):

$$P[mJ/s] = \alpha_u * th_u + \alpha_d * th_d + \beta \quad (7)$$

$\alpha_u, \alpha_d$ and $\beta$ are parameters, their values are different from one RAT to another [24], $th_u$ and $th_d$ are uplink and downlink throughput.

## 4. PERFORMANCE EVALUATION

In this section, we evaluate the performance of the proposed algorithm and we compare it with the MADM methods described in section II based on the input data from Table 3. In this study, we consider three types of services, VoIP, video service.
The study is composed of two parts:
- In the first part, we compare our proposal with the legacy MADM methods in the normal case where no RATs disappear in the middle of the selection process.
- The second part is the case when one network disappears from the list of the available networks, this case allows us to prove that MADM methods suffer from the rank reversal phenomenon (this is a well-known situation), and we will see if this problem occur or not in our proposal.

The following matrix represented by Table 2 represents the data input matrix where the values are randomly generated from Table 2; this matrix is used in our tests.

*Table 2*: The input matrix

|      | Bandwidth | Delay  | PLR  | Energy | Cost |
|------|-----------|--------|------|--------|------|
| N (0)| 1.730     | 105.85 | 7.94 | 1.00   | 0.2  |
| N (1)| 5.076     | 134.88 | 6.70 | 2.6    | 0.2  |
| N (2)| 6.849     | 43.98  | 2.84 | 6.26   | 1    |
| N (3)| 6.329     | 32.15  | 3.05 | 5.86   | 1    |
| N (4)| 66.66     | 95.15  | 6.32 | 12.78  | 0.4  |
| N (5)| 62.5      | 99.73  | 5.80 | 10.28  | 0.4  |

The weight values of each type of application VoIP, video services and the best effort applications are generated using the Eigenvector method using equation (8). We decided to use the Eigenvector method because it is already used in the AHP process. So, to be fair, we have decided to use the same method to get the weight vectors for all of the methods.

$$(mat - \gamma) * w = 0 \quad (8)$$

Where *mat* is the input matrix, $\gamma$ is the Eigen-value, $I$ is the identity matrix and $w$ is the associated eigenvector containing the weights values.
Table 3 contains the weights vector generated for each type of application.
We start with the first part of this study, the ordinary case, i.e. having all networks available.

*Table 3*: The weight vectors

|       | Bandwidth | Delay | PLR   | Energy | Cost  |
|-------|-----------|-------|-------|--------|-------|
| Wi-Fi | 0.047     | 0.486 | 0.371 | 0.047  | 0.047 |
| 3G    | 0.458     | 0.101 | 0.302 | 0.074  | 0.063 |
| LTE   | 0.299     | 0.146 | 0.146 | 0.108  | 0.299 |

In this study we are interested in the total ranking order of the networks and not only the network with the best rank because the best network is most likely becomes loaded quickly after a given time due to the connect request and then it will be unavailable, i.e. overloaded, so, it is necessary to get an optimal total ranking of all networks.

### A. Simulation 1: All RATs available

In this case, the networks in the covered area of the user do not disappear and remain available during the whole process.

- *VOIP scenario:*

The results in Table 4 concern the first case VoIP.

*Table 4*: Ranking results for VOIP

| Method | Rank  |       |       |       |       |       |
|--------|-------|-------|-------|-------|-------|-------|
| TOPSIS | N (3) | N (2) | N (0) | N (1) | N (5) | N (4) |
| AHP    | N (3) | N (2) | N (0) | N (4) | N (1) | N (5) |
| WPM    | N (3) | N (2) | N (0) | N (1) | N (5) | N (4) |
| SAW    | N (3) | N (2) | N (0) | N (1) | N (5) | N (4) |
| M-SAW  | N (3) | N (2) | N (4) | N (5) | N (0) | N (1) |

Based on results in Table 4, the methods TOPSIS, WPM and M-SAW give the same order for the first two networks. For the third position, our method M-SAW chooses the N (4) but TOPSIS, WPM and AHP choose N (0). So, we must compare the performance of N (4) and N (0) to see which method bring the right rank.

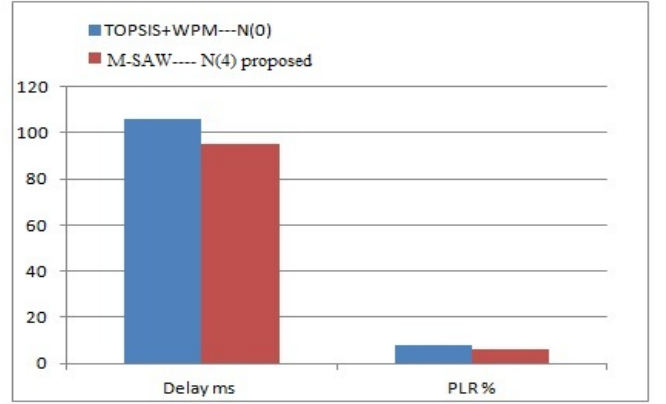

*Fig.2*: delay and PLR comparison between N (0) and N (4)

*Fig.3* shows that N (0) propose a delay of 105,85 and a PLR of 7,94. The N(4) has a delay of 95,15 and a PLR of 6,32. So, N (4) is better than N (0) and this means that our method brings the right rank order and it outperforms the TOPSIS, WPM, and AHP.
In the case VoIP, we showed that the proposed method M-SAW gives the best ranking order compared to the MADM methods, Table4 and *Fig.2*.

- *Video scenario*

The results in Table 4 concern the second case video service, using the matrix in Table 2.
From Table 5, the SAW method has always a wrong ranking order. For the other methods, TOPSIS chooses N (5) while WPM and M-SAW choose N (4).

**Table 5**: *Ranking results for Video service*

| Method | Rank | | | | | |
|---|---|---|---|---|---|---|
| TOPSIS | N (5) | N (4) | N (2) | N (3) | N (1) | N (0) |
| AHP | N (2) | N (3) | N (5) | N (4) | N (1) | N (0) |
| WPM | N (4) | N (5) | N (3) | N (2) | N (1) | N (0) |
| SAW | N (4) | N (2) | N (0) | N (3) | N (5) | N (1) |
| M-SAW | N (4) | N (2) | N (3) | N (5) | N (1) | N (0) |

From the Table 5, M-SAW and WPM select the N (4) and TOPSIS chooses N (5) as the best network.

In *Fig*.4, N (4) has higher bandwidth and lower delay. The N (5) has a better PLR, but in this case interactive scenario, the importance is given to the bandwidth and delay, so the best choice is the N (4).

Now, for the second place, WPM selects the N (5) and M-SAW select N (2). N (5) has a bandwidth of 62.5 and delay of 99,73 and a PLR of 5,80. N (2) has a bandwidth of 6, 85 and delay time of 43, 98 and a PLR of 2,84.

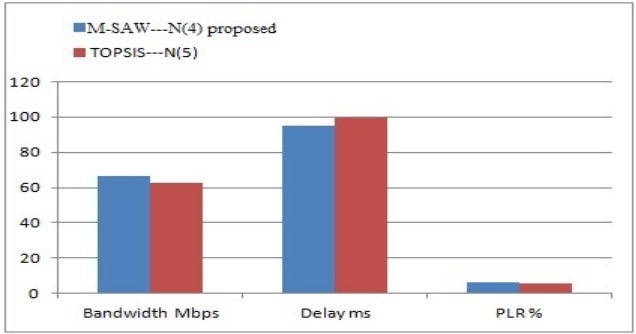

**Fig.3**: *Comparison between N (5) and N (4)*

Considering these values, we see that the N (5) has a larger bandwidth, but the bandwidth of N (2) is also good and can ensure the video service. In this case, we use the property of giving the user the minimum value of bandwidth that satisfies the application's requirement, this means that the application has some requirements, these requirements must be satisfied. Once the network satisfies these requirements it is considered as acceptable and the user can choose it. For the other parameters (delay and PLR), N (2) is very good compared to N (5), see Fig.4.

Based on *Fig*.4, our method M-SAW brings the best choices with this service type.

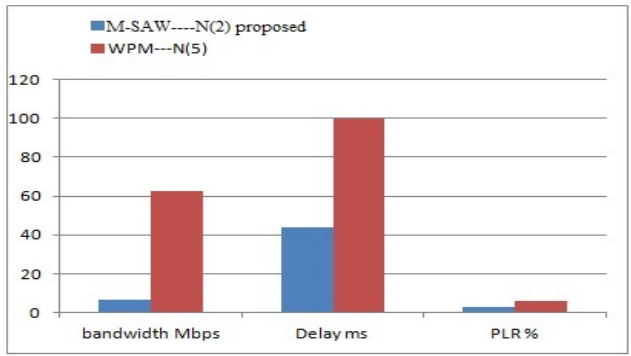

**Fig.4**: *Comparison between N(2) and N(5)*

In this subsection, we made a comparative study between our algorithm M-SAW and the MADM algorithms, our proposed algorithm gives the correct order of networks compared to the others algorithms in every case. This comparison showed us too that the TOPSIS methods had the most suitable ranking order among the legacy MADM algorithms.

The next step now consists of testing our function in the case when one RAT is added or removed, this case present what we call the ranking abnormality when using the legacy MADM. The following simulation will show us, whether our approach avoids this problem or it falls in it too.

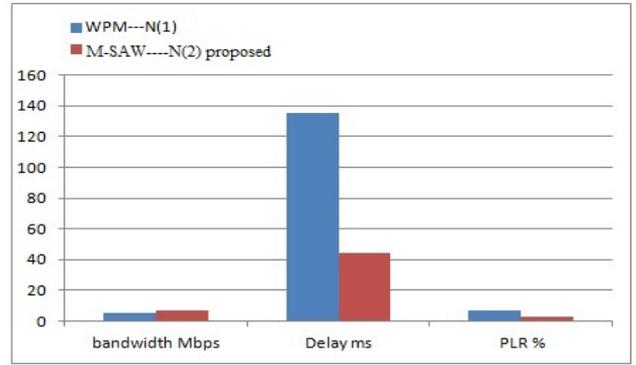

**Fig.5:** *Comparison between N (2) and N (1)*

B. *Simulation 2: The rank reversal case*

Now, we investigate the case that one network disappears in the middle of the ranking process, this case allows us to study the rank reversal phenomenon.

So, we decided to eliminate one network, for example, the network N (4) from the Table 2, and we repeat the simulation by ranking the remaining networks in the case of VoIP service, the results are shown in Table 6.

**Table 6:** *Ranking order in the case of disappearing of one RAT*

| Method | Rank | | | | |
|---|---|---|---|---|---|
| TOPSIS | N (5) | N (3) | N (2) | N (0) | N (1) |
| AHP | N (0) | N (1) | N (5) | N (3) | N (2) |
| WPM | N (5) | N (3) | N (2) | N (0) | N (1) |
| SAW | N (1) | N (3) | N (2) | N (0) | N (5) |
| M-SAW | N (3) | N (2) | N (5) | N (0) | N (1) |

Results in Table 6 show two information:
First, all the MADM (TOPSIS, SAW, WPM and AHP) methods suffer from the rank reversal phenomenon because the ranking results presented in Table 8 are different compared to those shown in Table 5, this result confirms the affirmation of the authors in [16] that all MADM methods suffer from the rank reversal phenomenon.

The second information is that our method M-SAW does not present the phenomenon of reversal's ranking when a network disappears; indeed, our proposed method just remove the disappeared alternative while keeping the other alternative's rank unchanged. This is a good result of our approach. Indeed, each network receives a local income equal to the mathematical multiplication between 'α' minus the rank value for this network and the weight value of the criterion. This new formulation of the weight allows us to control the income of each alternative according to its rank for the whole criteria.

To summarise, MADM methods give a ranking order that this is not always the optimal one; it is possible that one of these methods gives us the best-ranked network, so the user connects to this network. But, it is not always possible to connect to the best network given the number of users that select this best network, because, it will be easily loaded and then unreachable. To solve this problem we propose our algorithm that is a modified version of the SAW method. The goal of the algorithm is to find the most optimal total rank of all the available networks and not only the best network from a list of networks.

The second advantage of this algorithm is that it behaves well in the normal case when all the network remain available and in the case when one network disappears, in this case, the MADM methods present the problem of rank reversal.

## II. Conclusion

In the aim to find the best network at each instant, the idea was to rank the existing networks to get the total optimal ranking order; in this case, the operator switches the users with the best network available in the ranked list of networks. In this paper, we present our approach named modified-SAW, the objective function is based on the relative ranking order of each alternative for each criterion at each round of the process, and this basic idea allows us to get a greedy algorithm that gives good results. Results show that our proposed approach outperforms the existing used methods in the normal case, i.e. when all networks are available. Another test is done where one of the networks disappears, the simulation shows that all the MADM methods present the rank reversal phenomenon, our proposed algorithm overcomes this phenomenon and stays coherent and brings the same ranking order with eliminating the disappeared network.